# 4MOST Consortium Survey 5: eROSITA Galaxy Cluster Redshift Survey

Alexis Finoguenov[1,2]
Andrea Merloni[1]
Johan Comparat[1]
Kirpal Nandra[1]
Mara Salvato[1]
Elmo Tempel[3,4]
Anand Raichoor[5]
Johan Richard[6]
Jean-Paul Kneib[5]
Annalisa Pillepich[7]
Martin Sahlén[8]
Paola Popesso[9]
Peder Norberg[10]
Richard McMahon[11]
and the 4MOST collaboration

[1] Max-Planck-Institut für extraterrestrische Physik, Garching, Germany
[2] University of Helsinki, Finland
[3] Tartu Observatory, University of Tartu, Estonia
[4] Leibniz-Institut für Astrophysik Potsdam (AIP), Germany
[5] Laboratoire d'astrophysique, École Polytechnique Fédérale de Lausanne, Switzerland
[6] Centre de Recherche Astrophysique de Lyon, France
[7] Max-Planck-Institut für Astronomie, Heidelberg, Germany
[8] Department of Physics and Astronomy, Uppsala universitet, Sweden
[9] Physics Department, Technische Universität München, Germany
[10] Department of Physics, Durham University, UK
[11] Institute of Astronomy, University of Cambridge, UK

Groups and clusters of galaxies are a current focus of astronomical research owing to their role in determining the environmental effects on galaxies and the constraints they provide to cosmology. The eROSITA X-ray telescope on board the Spectrum Roentgen Gamma observatory will be launched in 2019 and will have completed eight scans of the full sky when 4MOST starts operating. The experiment will detect groups and clusters of galaxies through X-ray emission from the hot intergalactic medium. The purpose of the 4MOST eROSITA Galaxy Cluster Redshift Survey is to provide spectroscopic redshifts of the optical counterparts to the X-ray emission from 40 000 groups and clusters of galaxies so as to perform dynamical estimates of the total mass and to measure the properties of the member galaxies. The survey aims to obtain precise redshift measurements of the photometrically identified brightest cluster galaxies at redshift $z > 0.7$. At lower redshifts ($z < 0.7$) the programme aims to sample over 15 member galaxies per cluster and enable dynamical mass measurements to calibrate the clusters for cosmological experiments. At $z < 0.2$, eROSITA will also detect X-ray emission from galaxy groups and filaments. 4MOST spectroscopic data from the survey will be used for optical identification of galaxy groups down to eROSITA's mass detection limits of $10^{13}\,M_\odot$, as well as the detection of the largest filaments for pioneering studies of their X-ray emission.

## Scientific context

The experimental measurement of the precise values of the cosmological parameters is the accepted way to progress the field of cosmology. These observational measurements include: measuring the fluctuations of the cosmic microwave background, which describe the inhomogeneities of the Universe at a redshift of 1100; geometric tests of the expansion of the Universe (using Supernovae and Baryonic Acoustic Oscillations); and the growth of large-scale structure throughout cosmic time. As outlined in the Dark Energy Task Force report (Albrecht et al., 2006), no single method can constrain the cosmological parameters precisely and a combination of methods is required.

The performance of a particular survey is judged by its ability to constrain the dark energy equation of state. The 4MOST spectroscopic survey will provide the validation of the cluster catalogue produced by the eROSITA survey, which tests cosmology through the growth of structure as reflected in the mass function of galaxy clusters, and it belongs to the highest tier of dark energy experiments. The 4MOST eROSITA Galaxy Cluster Redshift Survey is the only currently planned large-scale programme of this kind. Other redshift survey experiments, such as the Dark Energy Spectroscopic Instrument (DESI) and the 4MOST Cosmology Redshift Survey, provide geometrical cosmological constraints, which are complementary; together they can constrain a larger variety of cosmological models.

Spectroscopic identification of galaxy clusters has been a cornerstone of all galaxy cluster surveys. Recently, this field has expanded due to developments on measurements of dynamical and caustic mass and the availability of precise memberships of cluster galaxies. In the last few years, a number of large-area galaxy cluster surveys have been carried out. The pioneering work on this has been done within the Sloan Digital Sky Survey SDSS-IV survey (Clerc et al., 2016), with a goal to follow up on 4000 clusters. A much larger demand on cluster follow-up is set by the eROSITA survey, and the goal of the 4MOST cluster survey is to follow up as many as 40 000 groups and clusters of galaxies, the precise details depending on the performance of the eROSITA survey. In addition to the larger number of systems, eROSITA will detect clusters to higher redshifts than before, reaching beyond $z = 1$, which requires a deeper optical survey than SDSS-IV. The survey will constrain the physics of the warm baryons, which in turn traces the evolution of the cosmic feedback. In addition to the properties of warm intergalactic gas in groups and clusters, X-ray emission traces the state of the warm gas in the filaments, which completes the picture of warm baryons in the Universe and complements similar studies using absorption techniques (Nicastro et al., 2018).

## Specific scientific goals

The primary scientific goal of the survey is to provide a spectroscopic study of 40 000 groups and clusters of galaxies within the German eROSITA sky (for details of the eROSITA survey, see Merloni et al., 2012), reaching a redshift of 1.4 to maximise the expected cosmological performance of the survey (for details of the cosmological forecast, see Pillepich et al., 2018). The primary uses of the 4MOST spectroscopic observations performed for the eROSITA Galaxy Cluster Redshift Survey are:

1. To spectroscopically confirm the photometric counterpart of the X-ray emission by removing projection effects in photometric galaxy membership assignment for galaxy clusters (Clerc et al., 2016) and to provide the spectroscopic counterparts to the X-ray emission coming from the galaxy groups (halos with total mass below $10^{14}\,M_\odot$).

2. To obtain a precise distance estimate required to compute X-ray luminosity, which enters the mass function cosmological study as a mass proxy (Grandis et al., 2018).

3. To perform dynamical mass calibration (Capasso et al., 2019). The survey will provide dynamical mass as well as caustic mass measurements for 10 000 clusters of galaxies at redshifts $z < 0.6$ and total masses $> 10^{14}\,M_\odot$, which enables us to very accurately link the cluster observables to their total mass — critical for constraining the cosmology. A comparison of these measurements to the weak lensing mass estimates can further constrain the models of modified gravity (Wilcox et al., 2015).

4. To enable the analysis of clustering of groups and clusters, which have a different sensitivity in constraining cosmology through the large-scale structure growth estimates. Clustering analysis requires at least 1000 systems in order to detect significant signal. A sample of 40 000 groups and clusters allows us to sample the mass function with five mass bins and the redshift range with eight bins, which is required to break the degeneracy of the clustering amplitude between mass, redshift and cosmology (Pillepich et al., 2018).

5. To improve the link between the various baryonic phases of halos and improve the value of eROSITA cosmology by addressing the effects of baryons on the growth of structure (Bocquet et al., 2016).

6. To spectroscopically identify filaments to a redshift of 0.2, where a detection of the X-ray signal with eROSITA is expected. The study of the Warm Hot Intergalactic Medium (WHIM) in emission will be based on the combined 4MOST and eROSITA study, through a cross-correlation analysis between the position of X-ray photons, detected by eROSITA and the 2D (sky) projections of the filaments.

7. The eROSITA Galaxy Cluster Redshift Survey will coordinate with the 4MOST WAVES Survey, where the eROSITA sample will provide high-signal-to-noise (S/N) observations of rare systems not sampled within the WAVES Survey area; 4MOST spectra with S/N > 10 per Å are suitable for galaxy evolution science. Examples of these types of rare systems are galaxies in rare high-density environments in massive clusters and X-ray selected lowest mass galaxy groups that are selected in the Local Volume.

### Science requirements

The science requirements for the 4MOST eROSITA Galaxy Cluster Redshift Survey consist of: achieving highly complete sampling of target galaxies required for the galaxy group and filament searches; covering a large area to maximise the number of spectroscopically confirmed groups and clusters of galaxies in order to improve the cosmological constraints; obtaining uniform coverage of sufficiently large areas to perform the clustering analysis to obtain additional cosmological constraints; delivering accurate calibration of cluster mass using dynamical and caustic mass measurement; and achieving high S/N for a subsample selected for galaxy evolution studies.

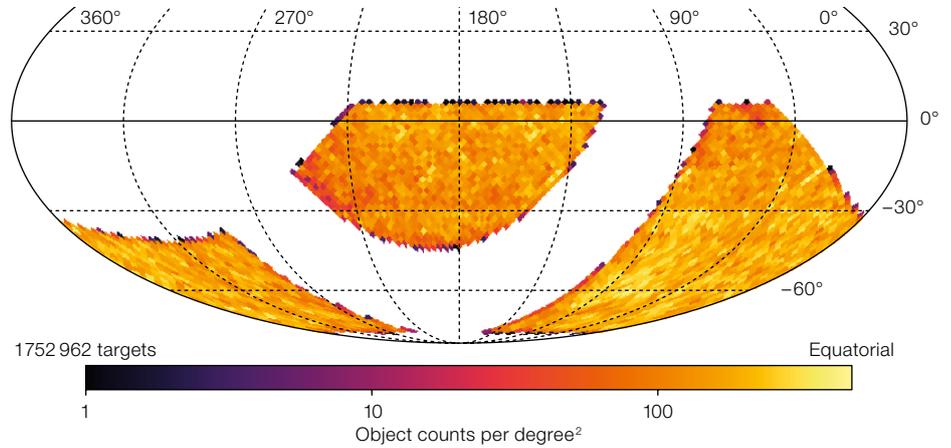

Figure 1. Target density of the 4MOST Cluster Survey, based on the mock catalogue tailored to the expectations of the eROSITA X-ray survey. About one third of the targets belong to the low-redshift ($z < 0.2$) survey.

### Target selection and survey area

The target selection for the eROSITA cluster identification is carried out using the position of the extended X-ray source and running the red sequence cluster finder on deep photometry provided by the Dark Energy Survey (DES), the Dark Energy Camera Legacy Survey (DECaLS), Pan-STARRS1, VST ATLAS and ongoing imaging surveys with the Dark Energy Camera (DECam) and VST (DeROSITAs/KABS). Altogether, these will cover 10 000 square degrees accessible by 4MOST (see Figure 1).

In addition, the project to search for filaments has a strong synergy with the 4MOST Cosmology Redshift Survey in terms of targets. Therefore this part of the 4MOST Cluster survey will only be carried out over the survey area in common with the Bright Galaxy sub-survey of the 4MOST Cosmology Redshift Survey (7500 square degrees). In order to estimate the resulting target density, we have produced a mock target catalogue by combining the expected eROSITA performance at cluster detection with the mocks based on the MultiDark simulation (Comparat et al., 2017 & in preparation). The low-redshift ($z < 0.2$) part of the survey has about one million targets, which are bright ($Ks < 18$ magnitudes) and require short exposures (10 minutes). Multi-member cluster identification at high redshift depends on the total exposure per field, with the trade-off being the highest redshift achievable within the total available time for the survey. The brightest cluster galaxies in each cluster are always considered for highest priority observations.

### Spectral success criteria and figure of merit

The spectral success criteria are the measurements of galaxy redshifts. The figure of merit (FoM) for the entire survey encompasses multiple components, including:
– Obtaining redshifts for a million targets to sample galaxy cluster members, the brightest cluster member in each cluster being the highest priority. An additional requirement is to determine redshifts for between 10 and 100 member galaxies within the virial radii for clusters with $z < 0.7$. This component requires 0.4 million fibre-hours in dark sky conditions to complete.
– A sufficiently large area to sample massive clusters, which are the most sensitive probes of the cosmology. The requirement (goal) is to survey 7500 (10 000) square degrees.
– Contiguous areas to enable the high-order statistical tests (two-point and three-point correlation functions). The requirement for the minimum size of each patch of contiguous area on the sky is 500 square degrees.
– A million bright targets to sufficiently sample low-redshift filaments and groups of galaxies. It requires 0.1 million fibre-hours in bright sky conditions to complete. As with the 4MOST WAVES Survey, this places a strong requirement for the survey completeness.

The FoM of the survey is a function of the ratio of the covered to total area $x$:

FoM = $0.5 + 0.5$ erf $((x - 0.75)/0.165)$

FoM = 0.5 (0.9); for an area of 7500 (9000) square degrees made of contiguous 500 square-degree patches.

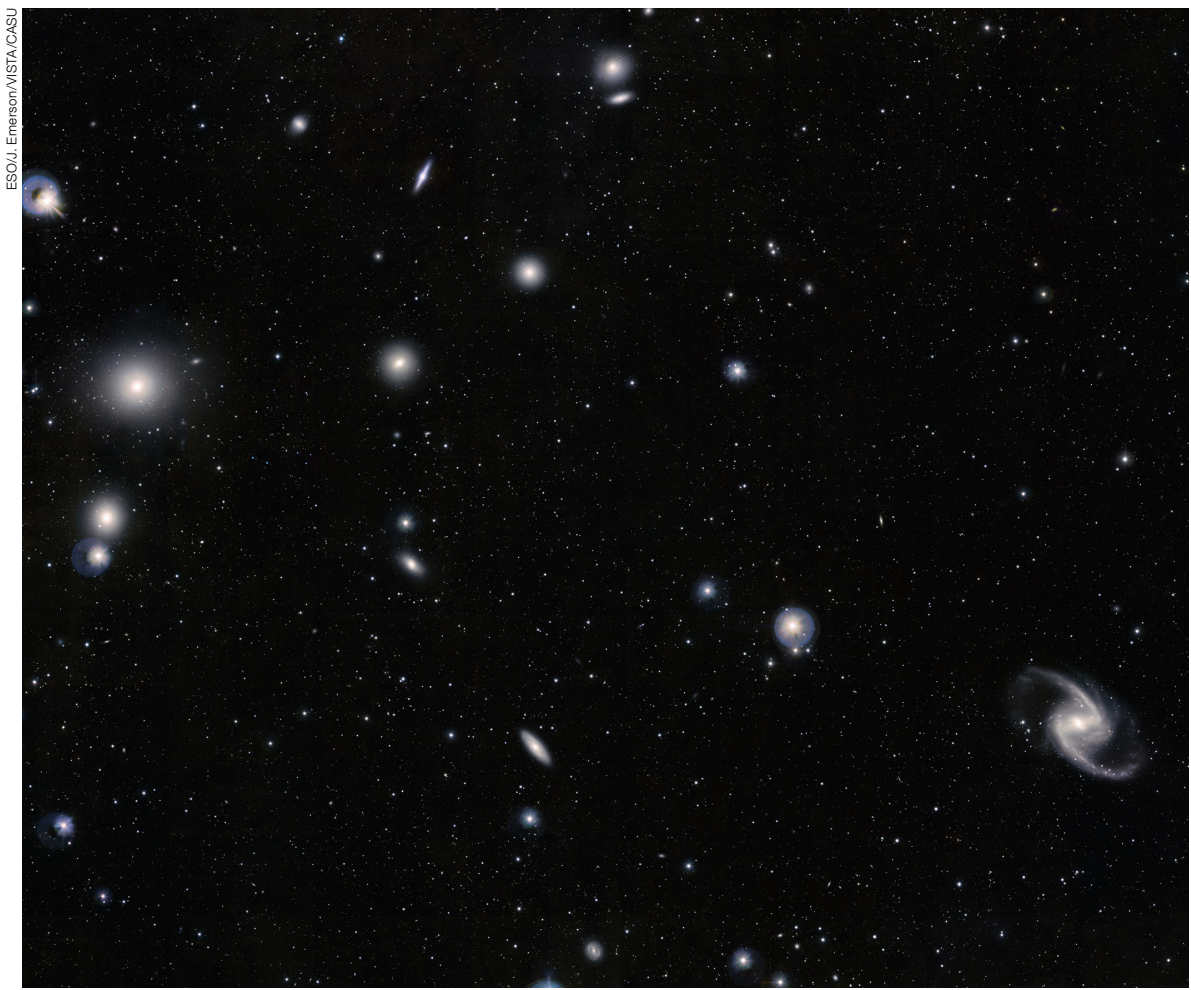

A VISTA image of the Fornax Galaxy Cluster, one of the closest clusters beyond the Local Group of galaxies.